\documentclass[10pt,letterpaper,twocolumn]{article}   
\usepackage{ol2}
\usepackage[draft]{hyperref}
\usepackage{amsmath}

\begin{document}

\twocolumn[
\title{Photon echoes generated by reversing magnetic\\ field gradients in a rubidium vapour}


\author{G.~H\'{e}tet$^1$, M.~Hosseini$^1$, B.~M.~Sparkes$^1$, D. Oblak$^{1,2}$, P.~K.~Lam$^{1*}$ and B.~C.~Buchler$^1$}
\address{$^1$ARC COE for Quantum-Atom Optics,  The Australian National University, Canberra ACT 0200, Australia\\
$^2$Danish National Research Council Centre for Quantum Optics (QUANTOP), Niels Bohr Institute, University of Copenhagen, Blegdamsvej 17, DK-2100 Copenhagen \O, Denmark}
\address{$^*$Corresponding author: ping.lam@anu.edu.au}

\begin{abstract} 
We propose a photon echo quantum memory scheme using detuned Raman coupling to long lived ground states. In contrast to previous 3-level schemes based on controlled reversible inhomogeneous broadening that use sequences of $\pi$-pulses, the scheme does not require accurate control of the coupling dynamics to the ground states.
We present a proof of principle experimental realisation of our proposal using rubidium atoms in a warm vapour cell.  The Raman resonance line is broadened using a magnetic field that varies linearly along the direction of light propagation.  Inverting the magnetic field gradient rephases the atomic dipoles and re-emits the light pulse in the forward direction.  \end{abstract}

] 
A robust photonic quantum information network requires a means of storing and retrieving light on demand.
 For ensemble systems, storage schemes using Electromagnetically Induced Transparency (EIT) \cite{polariton,hau,Gor,Novikova,EIT-CV} and Raman coupling \cite{Nunn,Koz,Dantan}  show great potential.  Another way of storing quantum information relies on photon echoes \cite{hartmann,Moss} generated using controlled reversible inhomogeneous broadening (CRIB) \cite{mois01,Nils05,krau06,sang06}. It was recently shown that with a linear atomic frequency gradient introduced along the axis of light propagation, it is possible to obtain an efficient photon echo in the forward direction by simply reversing the sign of the gradient  \cite{HetetPRL}. Experimental realisations of this Gradient Echo Memory (GEM) were performed using two-level atoms in a solid state system \cite{HetetPRL}. An advantage of GEM, and CRIB in general, is that a-priori knowledge of the temporal shape of the input pulse is not required for optimum storage \cite{Novikova,GEMpolariton}.
In this paper, we propose a three-level GEM scheme, using far off-resonance Raman coupling to ground states. We also present a proof of principle experiment in warm rubidium vapour using a spatially varying Zeeman shift.

\begin{figure}[t!]
  \centerline{\includegraphics[width=\columnwidth]{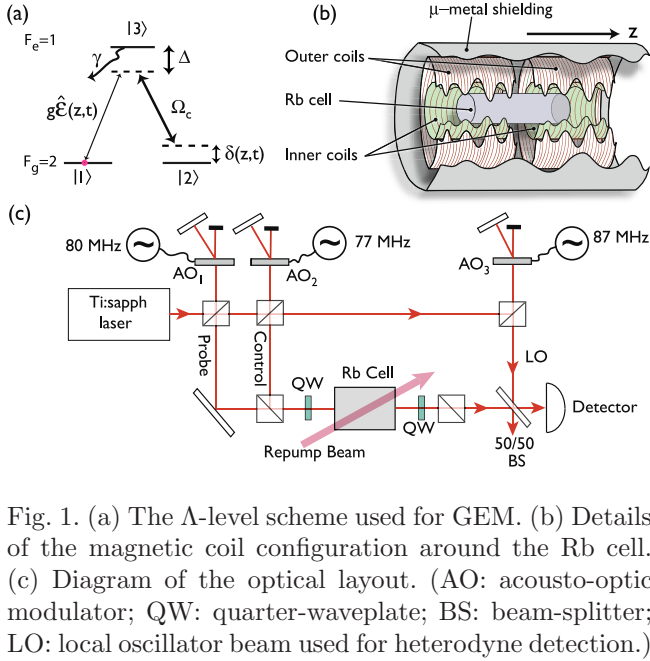}}
  \caption{(a) The $\Lambda$-level scheme used for GEM. (b) Details of the magnetic coil configuration around the Rb cell. (c) Diagram of the optical layout. (AO: acousto-optic modulator; QW: quarter-waveplate; BS: beam-splitter; LO:  local oscillator beam used for heterodyne detection.)}
  \label{setup}
  \end{figure}

We begin by demonstrating that Raman-coupled ground states can play an equivalent role to the two-level atom considered in previous GEM work \cite{HetetPRL} in the far detuned and adiabatic limits.  A three-level system is depicted Fig.~\ref{setup}(a) with a one-photon detuning $\Delta$, a two-photon detuning $\delta(z,t)$ that can be varied in time and be made linear in space, a classical control beam $\Omega_c$ and a weak quantum field $\hat{\mathcal{E}}$ that we wish to store.  The interaction Hamiltonian of the three-level system is
\begin{equation}\label{8H}
\hat{\mathcal{H}}= \hbar  (\Delta\hat\sigma_{33}+\delta(z,t)\hat{\sigma}_{22}+g\hat{\mathcal{E}}^{\dagger} \hat{\sigma}_{13} + \Omega_c^{*}  \hat{\sigma}_{23} +h.c )
\end{equation}
where $\hat{\sigma}_{ii}$ refers to the atomic population in the state $|i\rangle$, $\hat{\sigma}_{ij}$ is the atomic coherence of the transition $|i\rangle\rightarrow|j\rangle$ and $g$ is the atom-light coupling strength for the 1$\leftrightarrow$3 transition.
Assuming that all the population is in the ground state $|1\rangle$ initially, and that the probe is weak ($\hat{\sigma}_{11}\simeq1$), from the Heisenberg-Langevin equations, in a moving frame at the speed of light, we find
\begin{eqnarray}\label{8MBE}
\dot{\hat{\sigma}}_{13} & = & -(\gamma+\gamma_0/2+i\Delta)\hat{\sigma}_{13} + i g
\hat{\mathcal{E}}+ i\Omega_c \hat{\sigma}_{12} + \hat{F}_{13}. \nonumber \\
\dot{\hat{\sigma}}_{12} &=&  -(\gamma_{0}+i  \eta(t) z) \hat{\sigma}_{12}+ i \Omega_c^{\ast}
\hat{\sigma}_{13} + \hat{F}_{12} \label{max}\\
\frac{\partial}{\partial z}\hat{\mathcal{E}}&=& i g N \hat{\sigma}_{13} \nonumber ,
\end{eqnarray}
where $N$ is the linear atomic density and $\delta(z,t)=\eta(t) z $, which is linear in $z$. The Langevin operators $\hat{F}_{13}$ and $\hat{F}_{12}$ account for noise 
coming from spontaneous emission ($\gamma$) and ground state decoherence ($\gamma_0$) respectively.  It has been shown that these Langevin terms do not generate noise beyond that required to preserve the canonical commutation relations of the operators \cite{EIT-CV,Gor}.
To simplify the equations, we adiabatically eliminate fast excited state fluctuations by assuming $1/T \ll \gamma$, where T is the fastest time-scale of the system. We also assume large detuning compared to the spontaneous emission rate ($\Delta\gg\gamma$). In atomic ensembles with optical density $d=g^{2} N/\gamma$, $1/dT\ll \gamma $ and $\Delta\gg\gamma d$ are sufficient conditions when the optical density is larger than unity \cite{Gor}.  Equations~(\ref{max}) then become
\begin{eqnarray}
\dot{\hat{\sigma}}_{12} &=&  -\left [ \gamma_0+i  \eta(t) z \right ] \hat{\sigma}_{12}+i g'\hat{\mathcal{E}}  \nonumber \\
\frac{\partial}{\partial z}\hat{\mathcal{E}}&=& \phantom{-} ig' N\hat{\sigma}_{12}, \label{2lev}
\end{eqnarray}
where $g'= g \Omega_c/\Delta$. The reference frame has been normalised to the effective refractive index $igN/\Delta$ and the coupling beam frequency has been chosen to match the light shift $|\Omega_c|^2/\Delta$. Equations~(\ref{2lev}) are equivalent to the two-level memory presented in Ref. \cite{HetetPRL} where it was shown that reversing the detuning slope ($ \eta \rightarrow -\eta$) after the input light absorption causes an echo to leave the medium in the forward direction.  The echo was also shown to be an ideal time-reversed copy of the input \cite{GEMpolariton}.  In the case of the two-level system presented in \cite{HetetPRL}, there is a trade-off between long excited state lifetime and large optical depth, since both parameters are governed by the atom-light coupling $g$. By introducing a third level, the effective atom-light coupling ($g'$) and decoherence rate ($\gamma_{0}$) have been decoupled allowing, in principle, longer storage at higher optical depth. In theory, like other three-level memories, the storage time is limited by the ground state decay $\gamma_0$.  Provided the optical depth ($g'N/\eta$) is much larger than unity and the bandwidth of the storage medium ($\eta \times$ length of storage medium)  is larger than the pulse bandwidth,  the efficiency can be close to 100\% for storage times less than the ground state decay time.

\small
  \normalsize

The key advantage of this scheme, over previously proposed quantum memories using CRIB, is that the probe information is directly coupled to the ground states by the control field.  In principle, the dynamics of the control beam do not play a critical role in the memory process, and one can choose whether to leave the control beam on, or off, during the storage phase of the experiment. In other schemes, carefully timed pulses are required to efficiently map the probe into the ground state coherence.

Previous GEM experiments have used Stark shifts to prepare an atomic  frequency gradient of ions in a solid state material \cite{HetetPRL}. Our present experiment uses rubidium that has no linear Stark effect, hence a spatially varying magnetic field was employed to prepare a Zeeman gradient.  An arrangement of 4 coils, shown in Fig.~\ref{setup}(b), was used to generate the required magnetic fields.  The outer coils were used to generate a B-field of 3.4~G at the centre of the gas cell and a gradient in the $z$-direction that could be tuned to accommodate the spectral width of the input pulse.  The inner coils could be switched to reverse the sign of the B-field gradient. 

A schematic of the optical set-up is shown in Fig.~\ref{setup}(c). The Ti:Sapph laser was red detuned by 600 MHz from the transition $F_{g}=2\rightarrow F_{e}=1$ of the $^{87}$Rb $D_1$ line. The frequency of the control and probe could be tuned to match the two-photon detuning introduced by the magnetic field offset (3.4 G) and the light shift (50~kHz). The vapour cell contained isotopically enhanced $^{87}$Rb and 5~Torr of helium buffer gas. The control and probe beams had diameters of 2 and 0.3 cm respectively. An external cavity diode laser was used to repump atoms from the $F_g=1$ to the $F_g=2$ hyperfine levels.

\begin{figure}[t!]
  \centerline{\includegraphics[width=\columnwidth]{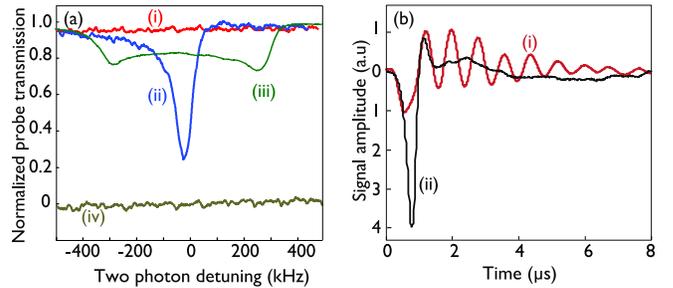}}
  \caption{(a) (i) Probe transmission without control beam.  (ii) Raman absorption without a magnetic field gradient. (iii) The absorption line broadened by the magnetic field gradient. (iv) Background level with probe light absent.  The parameters were: $\Delta=600$ MHz, control and probe beam powers of 40 mW and 1$\mu$W respectively, and a cell temperature of 60$^{\circ}$C. (b) The FID shown in (i) has an amplitude decay time of about 2.5$\mu s$.  With a magnetic field gradient the FID decay (ii) is much faster.}
\label{raman}
  \end{figure}

With the laser 600~MHz detuned from the Doppler-free transition frequency and no control beam, the incoherent absorption of probe light in the gas cell was 50\%, which is the level shown in trace~(i) of Fig.~\ref{raman}(a).  
To observe the Raman absorption line, the control beam was turned on and its frequency was scanned.  A Raman line of width 170~kHz was created that absorbs 75\% of the remaining probe light as shown in Trace~(ii).  Finally, Trace~(iii) shows the Raman line after Zeeman broadening is applied using the spatially varying B-field.

\small
\normalsize

After exciting the atoms participating in the Raman absorption with a pulse shorter than the inverse of the linewidth, the atomic macroscopic coherence will cause light to re-radiate via free induction decay (FID).  A long FID signal indicates a long macroscopic coherence time and a small ensemble linewidth.  To observe the FID, we used heterodyne detection and excited the atoms using a weak pulse of 250~ns.  In Fig.~\ref{raman}(b), Trace (i) shows the FID observed without applying a broadening magnetic field. To allow a precise measurement of the amplitude decay, the beat signal was mixed down with a frequency offset of 1 MHz. Oscillations of the FID amplitude at 1 MHz are observed for about 2.5 $\mu$s. The decay time of the intensity of the signal is then about 1.2 $\mu$s, consistent with the width of the unbroadened Raman feature ($1/ (2 \pi \times 170~{\rm kHz})\approx 1~\mu s$).  In the presence of a magnetic field gradient  the FID is very fast, as shown in Trace (ii).

\begin{figure}[t!]
  \centerline{\includegraphics[width=\columnwidth]{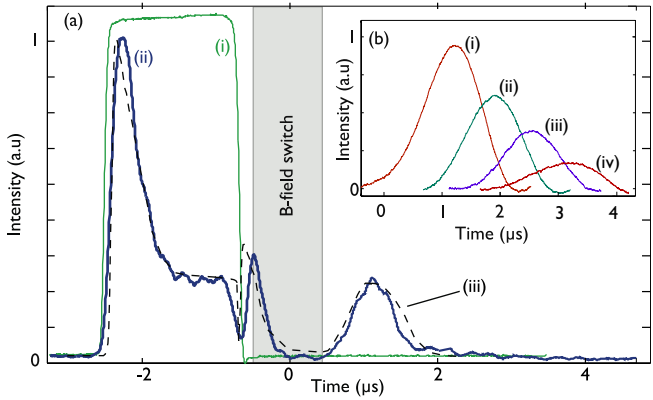}}
 \caption{(a) The input pulse measured without Raman absorption is shown in (i). Incoherent absorption in the gas cell was 50\% for this trace.  The absorption is further enhanced by the Raman beam, as shown by the left-hand side of (ii). After switching the B-field, an echo is generated as seen on the right-hand side of (ii).  Numerical modelling (iii) shows good agreement with the experiment.  (Parameters are $\gamma_{0}$=500~kHz, $\eta$=0.2~G/cm and a B-field switching time of 1~$\mu$s.) (b) As the magnetic field flipping is delayed in steps of 400ns from (i) to (iv), the pulse emerges later in steps of 800~ns.}
  \label{echos}
 \end{figure}

Having prepared the atomic ensemble and measured the FID, we found that by inverting the magnetic field slope, a pulse absorbed in the Zeeman shifted medium could be retrieved as an echo.
Fig.~\ref{echos}(a) shows the result of the experiment.  
  \small 
  \normalsize
Trace (i) shows the input pulse, around 1 $\mu$s long, measured by blocking the control field. Trace (ii) shows the light that was transmitted straight through the cell without being absorbed.  The shape of this transmitted pulse is a consequence of the low Fourier frequencies being coherently absorbed in the gas cell, while the higher frequencies are transmitted.  Further broadening of the Zeeman gradient leads to more uniform absorption of the Fourier spectrum at the expense of reducing the optical depth per unit bandwidth \cite{HetetPRL, GEMpolariton}.  After flipping the magnetic field slope at $t=0$~$\mu s$ using the inner coils, we retrieved part of the stored excitation.  About 30\% of the light that was coherently absorbed could be retrieved as an echo.  The absolute efficiency,  calculated as the area under the input and output pulses and accounting for the incoherent loss in the gas cell, was $\sim$1\%. The main features of the experiment are also reproduced by numerical simulations using the two-level atom model of Ref.~\cite{HetetPRL}. Trace (iii) shows the result of the simulations, using an atomic coherence time of 1$\mu$s, an optical depth of 0.4 and accounting for the finite switching rate of the magnetic field.  

By delaying the magnetic gradient flip in 400~ns steps, we observed a corresponding delay of $\sim$800~ns in the photon echo, as shown in  Fig.~\ref{echos}(b). The echo strength decays with a time constant of $\sim$1.5~$\mu$s, which is similar to the Raman linewidth and FID decay.
This decay rate is, however, faster than that found for EIT storage in the same gas cell, where a pulse could be stored for times of up to 20~$\mu s$.  Whereas in EIT only atoms with a small velocity spread in the longitudinal direction are addressed, in our experiment the detuned Raman transition acts over a wide range of longitudinal velocities.  This could lead to a ``blurred''  GEM spin wave and, since some of the faster atoms will have only a small one-photon detuning, increased spontaneous emission.  We therefore speculate that increased buffer gas pressures may lead to increased storage time.

In conclusion, we have proposed a scheme that extends the two-level atom GEM to three-level atoms with potentially longer coherence times. The simplicity of the proposal allowed us to demonstrate proof of principle experiments in a vapour cell.  We expect the performance to improve significantly with laser cooled atomic ensembles, cryogenic solid state systems, or vapour cells with optimised buffer gas pressure.

We would like to thank J.~J.~Longdell, M.~Sellars, J.~Cviklinski, A. ~Akulshin and M.~T.~L.~Hsu for enlightening discussions and experimental support.  We thank the Australian Research Council for financial support.

\end{document}